\documentclass[structabstract]{aa}  
\usepackage{graphicx}
\usepackage{txfonts}
\begin{document}

   \authorrunning{Groussin et al.}
   \titlerunning{The properties of asteroid (2867) Steins from Spitzer observations and OSIRIS shape reconstruction}

   \title{The properties of asteroid (2867) Steins from Spitzer Space Telescope observations and OSIRIS shape reconstruction}

   \author{O. Groussin
          \inst{1}
          \and
          P. Lamy
          \inst{1}
          \and
          S. Fornasier
          \inst{2}
          \and
          L. Jorda
          \inst{1}
          }

   \institute{Laboratoire d'Astrophysique de Marseille, UMR6110 CNRS/Universit\'e de Provence, 38 rue Fr\'ed\'eric Joliot Curie, 13388 Marseille Cedex 13, France\\
              \email{olivier.groussin@oamp}
         \and
             LESIA, Observatoire de Paris, 92195 Meudon Principal Cedex, France
             }

   \date{Received 02 November 2010; accepted --}

  \abstract
   {}
   {We report on the thermal properties and composition of asteroid (2867) Steins derived from an analysis of new Spitzer Space Telescope (SST) observations performed in March 2008, in addition to previously published SST observations performed in November 2005.}
   {We consider the three-dimensional shape model and photometric properties derived from OSIRIS images obtained during the flyby of the Rosetta spacecraft in September 2008, which we combine with a thermal model to properly interpret the observed SST thermal light curve and spectral energy distributions.}
   {We obtain a thermal inertia in the range 100$\pm$50~JK$^{-1}$m$^{-2}$s$^{-1/2}$ and a beaming factor (roughness) in the range 0.7-1.0. We confirm that the infrared emissivity of Steins is consistent with an enstatite composition. The November 2005 SST thermal light curve is most reliably interpreted by assuming inhomogeneities in the thermal properties of the surface, with two different regions of slightly different roughness, as observed on other small bodies, such as the nucleus of comet 9P/Tempel~1. 
Our results emphasize that the shape model is important to an accurate determination of the thermal inertia and roughness. 
Finally, we present temperature maps of Steins, as seen by Rosetta during its flyby, and discuss the interpretation of the observations performed by the VIRTIS and MIRO instruments.}
   {}


   \maketitle
%

\section{Introduction}

The Rosetta mission of the European Space Agency flew by asteroid (2867) Steins on 5 September 2008. Before this flyby, Lamy et al. (2008a) and Barucci et al. (2008) performed and analyzed Spitzer Space Telescope (hereafter SST) observations performed in November 2005, to determine the size, thermal properties, and infrared emissivity of the asteroid. At that time, the analysis was limited by its unknown size and shape, the uncertainties in the photometric properties, and the low signal-to-noise ratio (SNR) of the spectra used to derive the emissivity because its size was much smaller than anticipated. Since the Rosetta flyby, the size and shape (Jorda et al., 2011), and photometric properties (Spjuth et al., 2011) have been determined using OSIRIS images. To acquire spectra of higher SNR and better quality, we performed new SST observations of asteroid Steins in March 2008, before the Rosetta flyby. With this additional data, we can now perform a complete and improved analysis of the November 2005 and March 2008 SST data, to determine the infrared emissivity of Steins more accurately and refine its thermal properties (thermal inertia and roughness). These observations were not possible with Rosetta since none of its instruments cover the spectral range 5-40~$\mu$m.

\section{Spitzer Space Telescope observations}

In this article, we use two datasets of SST observations of asteroid Steins as summarized in Table~1. The first dataset was obtained on 22 November 2005, when Steins was 2.13~AU from the Sun, and 1.60~AU from SST, at a solar phase angle of 27.2$^{\circ}$, and is presented in detail in Lamy et al. (2008a). It consists of 14 spectra of Steins from 5 to 38~$\mu$m, taken over one rotation, from which Lamy et al. (2008a) derived its thermal light curve and its average infrared spectrum.

The second dataset was obtained on 30 March 2008, when Steins was 2.39~AU from the Sun, and 1.88~AU from the SST, at a solar phase angle of 23.6$^{\circ}$, and is presented for the first time in this article. Although Steins was slightly further away from the Sun and SST in March 2008 than in November 2005, the geometric conditions of the observations are close. We used the infrared spectrograph IRS in low-resolution mode to obtain five spectra of Steins in the wavelength range 5.2-38.0~$\mu$m. The short wavelength segments (SL2: 5.2-8.5~$\mu$m; SL1: 7.4-14.2~$\mu$m) were acquired with a 60~s ramp, while the long wavelength segments (LL2: 14.0-21.5~$\mu$m; LL1: 19.5-38.0~$\mu$m) were acquired with a 120~s ramp. We used the red peak-up camera at 22~$\mu$m to place Steins into the slit with high accuracy.

The March 2008 data were reduced using standard procedures developed for IRS, as described in detail in Lamy et al. (2008a). The data were first processed by the Spitzer Space Center using version S18.0.2 of the data reduction pipeline, to produce basic calibrated data (BCD). We then used the BCD to extract the spectrum of Steins using version 2.1.2 of the SPICE software with standard parameters. According to the SPICE software, the resulting uncertainty in the flux amounts to 3~\%. 

After data reduction, a mismatch between the different SL and LL modes from 6 to 16~\% was clearly visible, depending upon the mode. To correct for this effect, we used the following normalization factors for the different modes: 1.06 for SL2, 1.00 for SL1 (our reference), 1.15 for LL2, and 1.20 for LL1. These factors are accurate to about 5~\%, which increases the overall uncertainty in the flux to $\sim$6~\%. We believe that the mismatch between the different modes, taken at different times, results from the changing cross-section of the non-spherical rotating body. As we see in section~4, this effect can change the flux by 20~\% in a few tens of minutes.

An alternative explanation of the mismatch between the SL1 and LL2 modes could be the ``14 micron teardrop'', a known effect of the IRS instrument, which leads to excess emission at the end of the SL1 mode (13.2-15~$\mu$m). However, this effect is difficult to quantify (Spitzer Science Center, 2011, IRS Instrument Handbook version 3.0, http://ssc.spitzer.caltech.edu/irs/irsinstrumenthandbook/). In November 2005, the different parts of the spectrum, corresponding to the different modes (SL1, SL2, LL1, and LL2), were acquired almost instantaneously relative to the rotation period of the asteroid, and could all be smoothly connected without applying scaling factors. In particular, no jump was detected between SL1 and LL2, indicating a negligible ``teardrop'' in November 2005. Since we reobserved the same object in March 2008, we assume that the ``teardrop'' is likewise negligible. This assumption is discussed and justified in Section 5.

The average spectra of November 2005 and March 2008 are displayed in Fig.~1. The flux is lower in March 2008 than in November 2005, because of the larger heliocentric and SST-centric distances. However, thanks to the longer integration time, the SNR is higher in March 2008.

\begin{table*}
\caption[]{Spitzer Space Telescope observations of asteroid Steins.}
\label{observations}
\begin{tabular}{lcccccc}
\hline
\noalign{\smallskip}
Date (start-stop)	&$r_{\rm h}$	&$\Delta$	&$\alpha$	&$\lambda_1 -\lambda_2$	&Number		&Integration time\\
	&(AU)	&(AU)		&(degree)	&($\mu$m)		&of spectra	&(sec)	\\
\noalign{\smallskip}
\hline
\noalign{\smallskip}
22 Nov. 2005 UT 06:23-06:31	&2.13	&1.60	&27.2	&5.2-38.0		&1	&6 (SL mode) or 15 (LL mode)\\
22 Nov. 2005 UT 07:03-07:10	&2.13	&1.60	&27.2	&5.2-38.0		&1	&6 (SL mode) or 15 (LL mode)\\
22 Nov. 2005 UT 07:38-07:45	&2.13	&1.60	&27.2	&5.2-38.0		&1	&6 (SL mode) or 15 (LL mode)\\
22 Nov. 2005 UT 08:13-08:20	&2.13	&1.60	&27.2	&5.2-38.0		&1	&6 (SL mode) or 15 (LL mode)\\
22 Nov. 2005 UT 08:35-08:42	&2.13	&1.60	&27.2	&5.2-38.0		&1	&6 (SL mode) or 15 (LL mode)\\
22 Nov. 2005 UT 09:08-09:15	&2.13	&1.60	&27.2	&5.2-38.0		&1	&6 (SL mode) or 15 (LL mode)\\
22 Nov. 2005 UT 09:32-09:39	&2.13	&1.60	&27.2	&5.2-38.0		&1	&6 (SL mode) or 15 (LL mode)\\
22 Nov. 2005 UT 10:06-10:11	&2.13	&1.60	&27.2	&5.2-38.0		&1	&6 (SL mode) or 15 (LL mode)\\
22 Nov. 2005 UT 10:38-10:45	&2.13	&1.60	&27.2	&5.2-38.0		&1	&6 (SL mode) or 15 (LL mode)\\
22 Nov. 2005 UT 11:04-11:12	&2.13	&1.60	&27.2	&5.2-38.0		&1	&6 (SL mode) or 15 (LL mode)\\
22 Nov. 2005 UT 11:35-11:42	&2.13	&1.60	&27.2	&5.2-38.0		&1	&6 (SL mode) or 15 (LL mode)\\
22 Nov. 2005 UT 12:12-12:19	&2.13	&1.60	&27.2	&5.2-38.0		&1	&6 (SL mode) or 15 (LL mode)\\
22 Nov. 2005 UT 12:36-12:43	&2.13	&1.61	&27.2	&5.2-38.0		&1	&6 (SL mode) or 15 (LL mode)\\
22 Nov. 2005 UT 13:11-13:18	&2.13	&1.61	&27.2	&5.2-38.0		&1	&6 (SL mode) or 15 (LL mode)\\
\noalign{\smallskip}
30 Mar. 2008 UT 03:05-03:09	&2.39	&1.88	&23.6	&5.2-8.5 (SL2)		&5	&60\\
30 Mar. 2008 UT 03:22-03:28	&2.39	&1.88	&23.6	&7.4-14.2 (SL1)		&5	&60\\
30 Mar. 2008 UT 03:35-03:47	&2.39	&1.88	&23.6	&14.0-21.5 (LL2)	&5	&120\\
30 Mar. 2008 UT 04:00-04:13	&2.39	&1.88	&23.6	&19.5-38.0 (LL1)	&5	&120\\
\noalign{\smallskip}
\hline
\end{tabular}
\\$r_{\rm h}$: heliocentric distance. $\Delta$: distance from Spitzer. $\alpha$: solar phase angle. $\lambda_1 -\lambda_2$: spectral range.
\end{table*}

\begin{figure} [!h] 
\includegraphics[width=\linewidth]{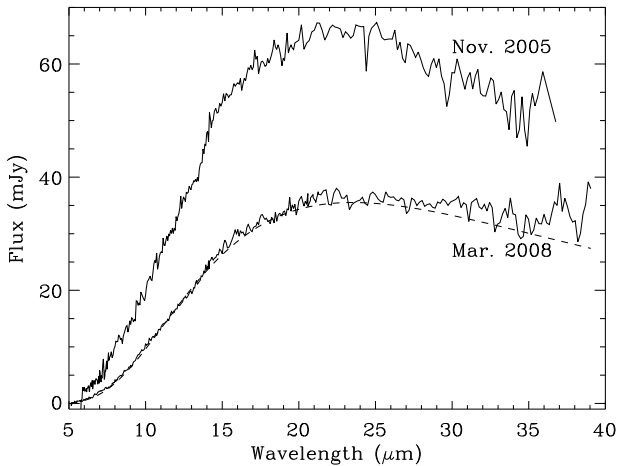}
\caption{November 2005 and March 2008 SST spectra of asteroid Steins, between 5 and 38~$\mu$m. Thanks to the longer integration time, the SNR is higher in March 2008. The dashed line represents the continuum used to derived the emissivity (Fig.~7), which corresponds to a black body at a temperature of 217~K.}
\end{figure}

\section{Nucleus model}

The interpretation of the infrared data requires a thermal model that decribes the energy balance on the surface between the flux received from the Sun, the re-radiated flux and the heat conduction into the asteroid. This thermal model is associated with a shape model that must be used to properly describe the thermal infrared light curve.

\subsection{Shape model}

We use the shape model of asteroid Steins constructed by Jorda et al. (2011). This model is based on several techniques involving non-resolved and resolved images. The portion of the asteroid imaged by Rosetta/ORISIS is determined by stereo-photo-clinometry, while the portion that is not imaged is constrained by the inversion of a large set of light curves (Lamy et al. 2008b). The overall dimensions of the model are 6.8~$\times$~5.7~$\times$~4.4~km as measured along the principal axis of inertia. We use a re-sampled version of the model of Jorda et al. (2011), composed of 1144 triangular facets. The pole orientation of the model is defined by R.A.=90.7~$^o$ and DEC=-62.1~$^o$, and the rotation period is 6.04681$\pm$0.00002~h (Lamy et al. 2008b). Fig.~2 illustrates the shape model of asteroid Steins as seen from the SST in November 2005 and March 2008.

\begin{figure*} [!t] 
\includegraphics[width=\linewidth]{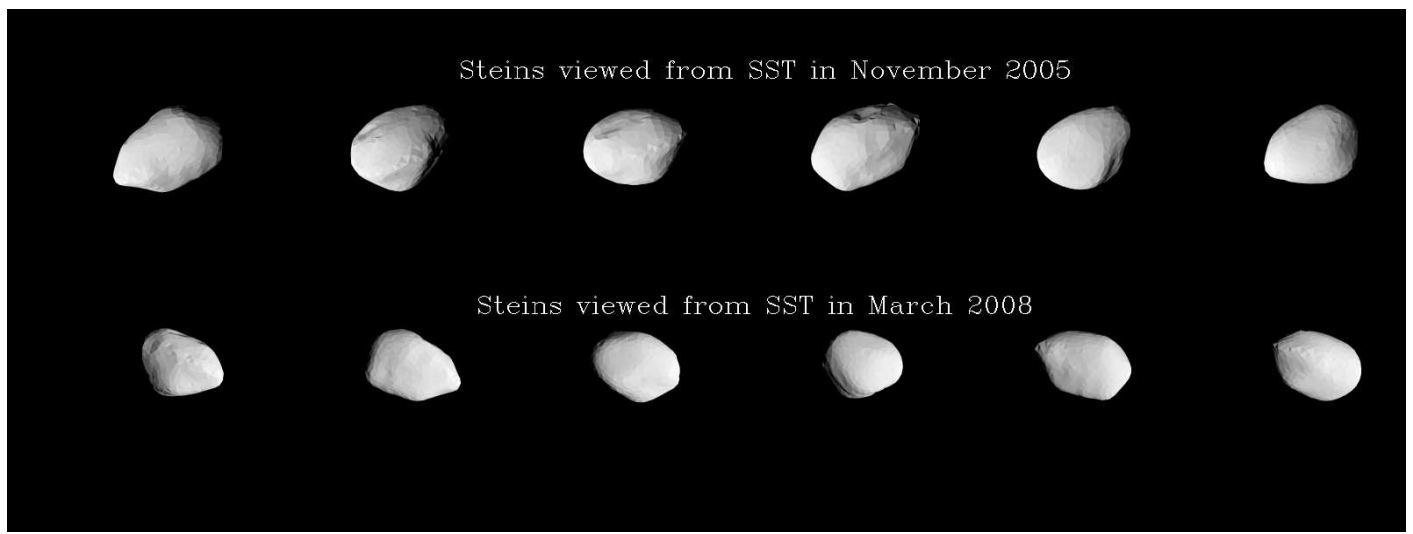}
\caption{The shape of asteroid Steins as seen from Spitzer Space Telescope in November 2005 (top) and March 2008 (bottom). Six positions are represented, covering one rotation period. In November 2005 the Sun was to the left, and on March 2005 to the right.}
\end{figure*}

\subsection{Thermal model}

We use the thermal model presented in Lamy et al. (2008a). The surface energy balance for each facet of the shape model is given by
\begin{equation} \label{model_equation}
(1-A(\lambda)) \frac{F_{sun}(\lambda)}{r_{\rm h}^2} v \cos(z) = 
\eta \epsilon \sigma T^4 + 
\kappa \frac{\partial T}{\partial x}\bigg|_{x=0}
\end{equation}
where $A(\lambda)$ is the Bond albedo as a function of wavelength $\lambda$, $F_{sun} (\lambda)$ [Wm$^{-2}$] is the solar flux as a function of wavelength, $r_{\rm h}$ (AU) is the heliocentric distance, $v$ is the illumination factor of the facet ($v=1$ if the facet is illuminated and $v=0$ if the facet is in shadow), $z$ is the zenithal angle of the facet, $\eta$ is the beaming factor introduced by Lebofsky and Spencer (1989), $\epsilon$ is the infrared emissivity, $T$ (K) is the surface temperature of the facet, $\kappa$ (Wm$^{-1}$K$^{-1}$) is the thermal conductivity, and $x$ measures the depth. As the asteroid rotates around its spin axis, the value of $z$ changes.

We solve Eq.~(1) using a method similar to that of Spencer et al. (1989), described in Groussin et al. (2004). It introduces the thermal inertia $I=\sqrt{\kappa \rho C}$, where $\rho$ (kg/m$^{3}$) is the bulk density, and $C$ (J/kg/K) is the specific heat capacity of the asteroid. We use a time step of 7~sec, which is small enough relative to the rotation period ($\sim$6.05 hr) to ensure relaxation of the numerical solution in 100 rotations. As a result, we obtain the temperature of each facet as a function of time, over one period of rotation. From this temperature profile, we calculate the infrared flux $F(\lambda)$ from each facet as a function of time using Eq.~(2)
\begin{equation}
F(\lambda)=\frac{\epsilon}{\Delta^2} B(\lambda,T) u \cos(w) dS
\end{equation}
where $\Delta$ (km) is the distance to the observer (SST in our case), $B(\lambda,T)$ is the Planck function, $u$ is the view factor between the facet and the observer ($u=1$ if the observer see the facet, $u=0$ otherwize), $w$ is the angle between the facet and the observer, and $dS$ is the surface area of the facet. The total flux received by the observer is the sum of the individual fluxes of all facets of the shape model. It is computed at different wavelengths to construct the spectral energy distribution.

\subsection{Parameters of the thermal model}

Our model has four free parameters: the infrared emissivity $\epsilon$, the Bond albedo $A(\lambda)$, the beaming factor $\eta$, and the thermal inertia $I$.

The infrared emissivity $\epsilon$ is assumed to be 0.95, the middle point of the interval 0.9-1.0 quoted in the literature. Since the interval is very small and the value near 1.0, this uncertainty has a negligible influence on the calculated thermal flux. 

The Bond albedo $A(\lambda)$ was estimated by Spjuth et al. (2011) using Rosetta OSIRIS images and we use their value. The Bond albedo is wavelength dependent and equals to 0.241 at 630~nm.

The beaming factor $\eta$ follows the strict definition given by Lagerros (1998) and therefore only reflects the influence of surface roughness. In theory, $\eta$ ranges from 0 (largest roughness) to 1 (flat surface), but in practise must be larger than 0.7 to avoid unrealistic roughness, with r.m.s. slopes exceeding 45~deg (Lagerros, 1998). In  this study, $\eta$ is unknown and derived from the observations.

For the thermal inertia, we consider six different values $I$=0, 10, 50, 100, 150, and 200~JK$^{-1}$m$^{-2}$s$^{-1/2}$, which cover and even extend the range determined by Lamy et al. (2008a) of 150$\pm$60~JK$^{-1}$m$^{-2}$s$^{-1/2}$. As we demonstrate later (section~4), larger values are incompatible with the observations.

Since $\epsilon$ and $A(\lambda)$ are well constrained, we can only vary the thermal inertia $I$ and the beaming factor $\eta$. These two parameters have a similar effect of changing the surface temperature distribution. They cannot be determined independently and to each value of $I$ corresponds a value for $\eta$.

\section{Analysis of the thermal light curve}

Using the above thermal and shape models, we generate a synthetic thermal light curve of asteroid Steins as seen from the SST in November 2005. For each value of the thermal inertia 0, 10, 50, 100, 150, and 200~JK$^{-1}$m$^{-2}$s$^{-1/2}$, we can fit the observations to determine the best value for $\eta$. The determination of $\eta$ and the phasing of the light curve are performed using a least squares technique. The results are presented Table~2. Owing to the 6\% uncertainty in the flux calibration, $\eta$ is accurate to $\pm$0.05.

\begin{table}
\begin{center}
\caption[]{Thermal inertia and beaming factor of asteroid Steins derived from the November 2005 SST data.}
\begin{tabular}{cc}
\hline
\noalign{\smallskip}
Thermal inertia $I$ &Beaming factor $\eta$\\
(JK$^{-1}$m$^{-2}$s$^{-1/2}$)	&\\
\noalign{\smallskip}
\hline
\noalign{\smallskip}
0	&1.05 $\pm$ 0.05\\
10	&1.04 $\pm$ 0.05\\
50	&0.94 $\pm$ 0.05\\
100	&0.79 $\pm$ 0.05\\
150	&0.69 $\pm$ 0.05\\
200	&0.61 $\pm$ 0.05\\
\noalign{\smallskip}
\hline
\end{tabular}
\normalsize
\end{center}
\end{table}

\begin{figure} [!t] 
\includegraphics[width=\linewidth]{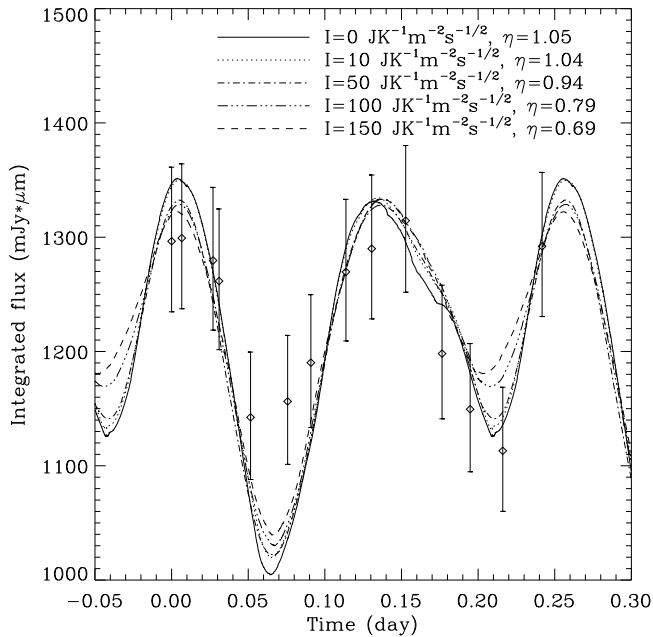}
\caption{Synthetic thermal light curves of asteroid Steins for different combinations of thermal inertia $I$ and beaming factor $\eta$ and the data obtained in November 2005 (diamonds).}
\end{figure}

Since we restrict the value of roughness ($\eta$) to the range 0.7-1.0 as described in section~3.3, the only possible values of the thermal inertia are 0-150~JK$^{-1}$m$^{-2}$s$^{-1/2}$, taking into account the $\pm$0.05 uncertainty in $\eta$. Values of $I$=200~JK$^{-1}$m$^{-2}$s$^{-1/2}$ and larger are incompatible with the observations as they imply that $\eta<0.7$, i.e., an unrealistic roughness.

Figure~3 illustrates the synthetic light curves of asteroid Steins, as seen by the SST on November 2005, for the acceptable values of thermal inertia between 0 and 150~JK$^{-1}$m$^{-2}$s$^{-1/2}$ of Table~2. To first order, the agreement is acceptable regardless of the thermal inertia, since all the synthetic light curves reproduce the observed double peaks, and are consistent with 12 out of the 14 data points at the 1$\sigma$ level. However, the first minima at $\sim$0.06~days is not well reproduced, the modeled fluxes being too low. This first minima corresponds to the part of Steins observed by OSIRIS and is therefore well defined, so that a shape effect seems excluded. We interpret this mismatch between the synthetic and observed thermal light curves as resulting from inhomogeneities in the thermal properties on the surface. 

Figure~4 illustrates the case where Steins has a thermal inertia of 100~JK$^{-1}$m$^{-2}$s$^{-1/2}$ and two different roughnesses: $\eta$=0.79 on about 3/4 of its surface and $\eta$=0.72 on the remaining 1/4. The transition between these two regions is smoothed out by using a combination of the two roughnesses. The second minimum is well reproduced with $\eta$=0.79, as in Fig.~3, but the first minimum is now also well reproduced by instead assuming $\eta$=0.72. The variation from $\eta$=0.79 to $\eta$=0.72 between two different regions is reasonable as it roughly corresponds to a small change in the surface r.m.s. slope of about 10$^{\circ}$. These variations have already been observed across the nucleus of comet 9P/Tempel~1 for example, there being an average r.m.s. slope of about 12-16$^{\circ}$ for the whole observed surface, but for specific regions slopes of 24-32$^{\circ}$ (Li et al., 2007; Spjuth et al., 2011). 

\begin{figure} [!t] 
\includegraphics[width=\linewidth]{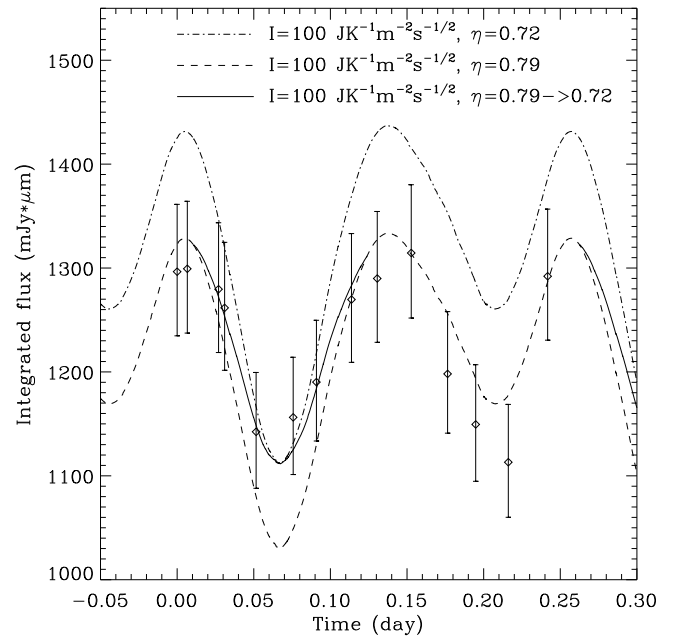}
\caption{Synthetic thermal light curves of asteroid Steins for a thermal inertia $I$ of 100~JK$^{-1}$m$^{-2}$s$^{-1/2}$ and two different values of $\eta$, i.e., 0.79 and 0.72. The solid line corresponds to the solution that combines two regions with these two different roughnesses.}
\end{figure}

Similar results are obtained for $I$=0~JK$^{-1}$m$^{-2}$s$^{-1/2}$ (respectively, 10, 50, and 150), where about 3/4 of the surface has $\eta$=1.05 (respectively, 1.04, 0.94, and 0.69) and the remaining 1/4 $\eta$=0.96 (respectively, 0.95, 0.85, and 0.62). In the latter case ($I$=150), the value of $\eta$=0.62 ($\pm$0.05) becomes slightly lower than the limit of 0.7 imposed above, but we prefer to be conservative and consider that this solution could also be valid.

In Fig.~3, each point of the light curve is derived from an observed spectral energy distribution (SED), displayed in Fig.~5. For the two extreme combinations (I=0, $\eta$=1.05) and (I=150, $\eta$=0.69) of Table~2, and using our shape model, we compute the synthetic SED corresponding to each observed SED. The results are illustrated in Fig.~5. Owing to the quasi-similarities of the corresponding curves, the two combinations (I, $\eta$) are possible and we cannot favor one to the other. We repeat the same work for the other intermediate combinations of Table~2 and find that they are also acceptable. To summarize, the data from November 2005 are compatible with a thermal inertia in the range 0-150~JK$^{-1}$m$^{-2}$s$^{-1/2}$.

In Lamy et al. (2008a), the shape model was not known with sufficient accuracy to perform such a detailed analysis and we used the mean of the fourteen individual SEDs (dispayed on Fig.~1), with a spherical shape model, thus resulting in slightly different results (I=150$\pm$60~JK$^{-1}$m$^{-2}$s$^{-1/2}$). This emphasizes the importance of using an accurate shape model to interpret infrared thermal light curves to properly differentiate the effects of the shape from that of the thermal parameters.

\begin{figure*} [!t] 
\includegraphics[width=\linewidth]{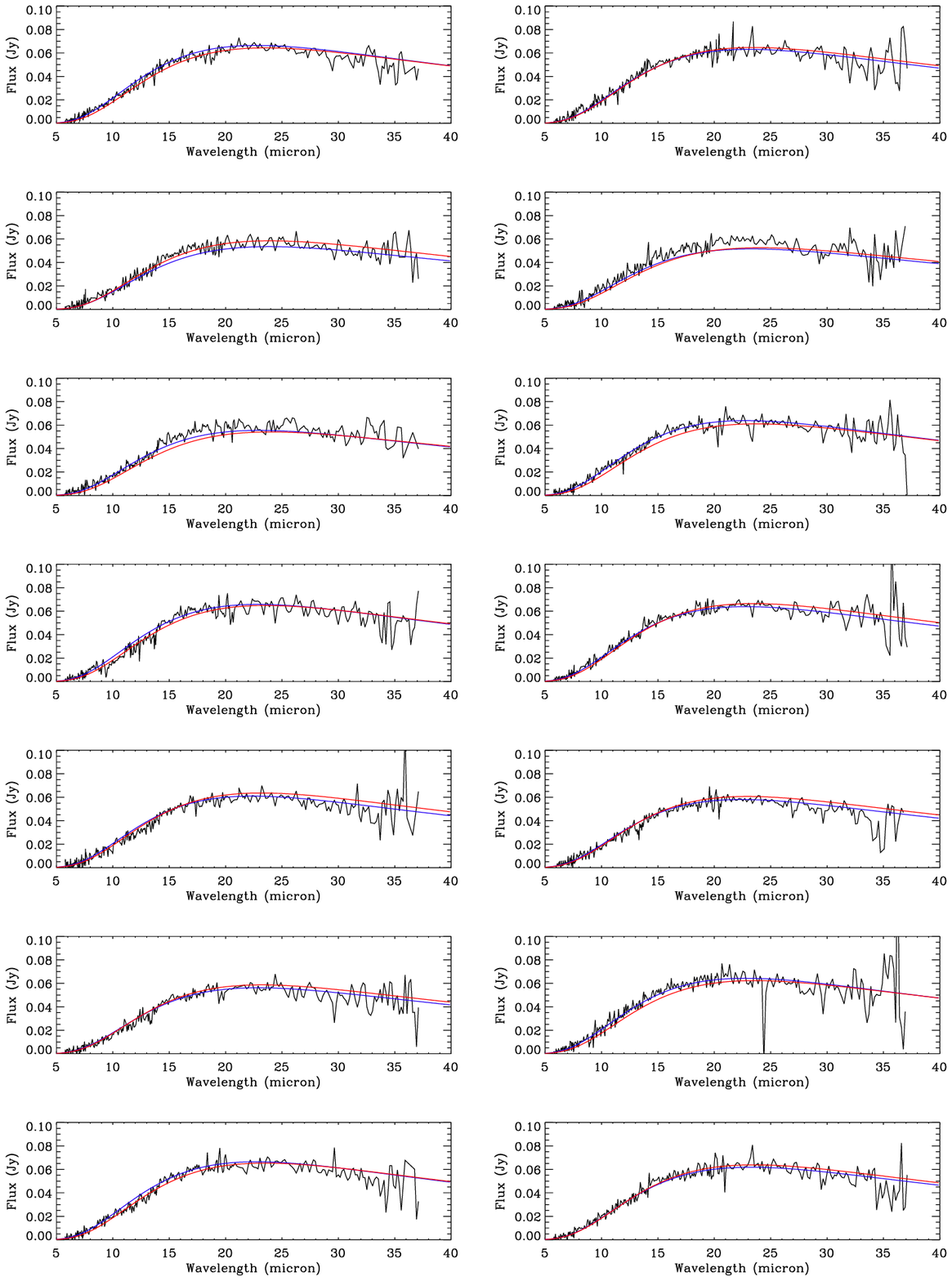}
\caption{Individual spectral energy distribution (SED) observed in November 2005, with their corresponding synthetic SED for the extreme combinations of thermal parameters : blue line I=0~JK$^{-1}$m$^{-2}$s$^{-1/2}$, $\eta$=1.05; red line I=150~JK$^{-1}$m$^{-2}$s$^{-1/2}$, $\eta$=0.69.}
\end{figure*}

Once we have determined the thermal parameters for November 2005, we can check their consistency with the March 2008 data. Figure~6 illustrates the synthetic thermal light curve of asteroid Steins as seen from the SST for the valid combinations ($0 \leq I \leq 150$~JK$^{-1}$m$^{-2}$s$^{-1/2}$) derived above (Table~2). The spectra from March 2008 were taken at different times in different modes (see Table~1), thus providing information on the flux variations due to the rotation of the asteroid. Its rotation period is not accurate enough to obtain the absolute phase relative to November 2005 and the phase shift was adjusted to match the March 2008 data. According to the normalization factors presented in section~2 for the different modes and as illustrated in Fig.~6, we most likely observed the maximum cross-section. The low thermal inertia of 0-10~JK$^{-1}$m$^{-2}$s$^{-1/2}$ are unambiguously excluded. Therefore, the thermal inertia of Steins must lie in the range 50-150~JK$^{-1}$m$^{-2}$s$^{-1/2}$, that is 100$\pm$50~JK$^{-1}$m$^{-2}$s$^{-1/2}$, in agreement with the March 2008 data.

\begin{figure} [!t] 
\includegraphics[width=\linewidth]{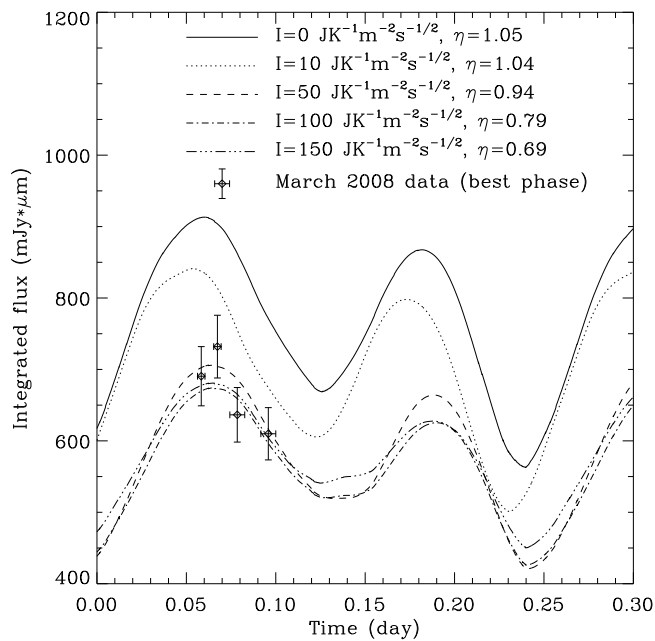}
\caption{Synthetic thermal light curves of asteroid Steins for different combinations of thermal inertia $I$ and beaming factor $\eta$ and the data obtained in March 2008 (diamonds). Each data point corresponds to the March 2008 integrated spectrum displayed in Fig.~1, normalized by the factors introduced in Section~2: 1.06 for SL2, 1.00 for SL1 (our reference), 1.15 for LL2, and 1.20 for LL1.}
\end{figure}

\section{Infrared emissivity}

The mid-infrared spectral domains contain features that are diagnostic of the surface composition, such as the Christiansen peaks (8-9.5~$\mu$m), Reststrahlen bands (9-12~$\mu$m and 14-25~$\mu$m), and Transparency features (11-13~$\mu$m). Using the November 2005 SST data, Barucci et al. (2008) detected these features on asteroid Steins, and by comparing the observed spectra with laboratory measurements, concluded that the thermal emissivity of Steins is similar to those of enstatite achondrite meteorites, which is consistent with Steins being an E-type asteroid.


Figure~7 illustrates the infrared emissivity of asteroid Steins derived from the March 2008 data, which is obtained by dividing the observed spectra by the continuum. In our case, the continuum is a black body with a color temperature of 217~K, as displayed in Fig.~1. The infrared emissivity of asteroid Steins obtained in March 2008 is compared to that of November 2005 and to laboratory measurements for enstatite and aubrite. The Christiansen (C), Reststrahlen (R), and Transparency (T) features are clearly visible in March 2008. The correspondence between both the C and R peaks and those of enstatite and aubrite is much better than for the November 2005 case, because of the higher SNR. Moreover, the broad emission at $\sim$15~$\mu$m is also more accurately reproduced. It reinforces the conclusion of Barucci et al. (2008) that Steins has an enstatite composition.

As explained in Section~2, we assume a negligible ``14 micron teardrop'' for the nominal spectrum of March 2008. To justify this assumption, we display in Fig.~7 the same spectrum, but assuming a non-negligible  ``teardrop'', i.e., ignoring the SL1 flux beyond 13.2~$\mu$m. In this case, the Christiansen and Reststrahlen peaks are still present, but the Transparency feature disappears, which contradicts the results of November 2005, when the ``teardrop'' was negligible. The consistency between the November 2005 and March 2008 emission spectra justifies our assumption.


\begin{figure} [!t] 
\includegraphics[width=\linewidth]{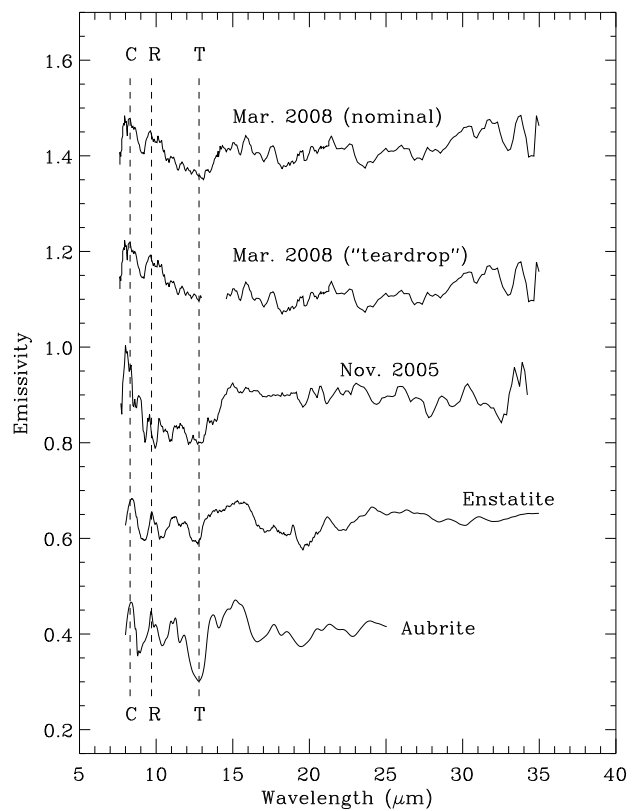}
\caption{Spectral variation in the infrared emissivity of asteroid Steins derived from the March 2008 observations, compared to that derived from the November 2005 observations and to two laboratory measurements for enstatite and aubrite (Barucci et al., 2008). In the case of the March 2008 data, two spectra are represented, the nominal spectrum and an alternative spectrum incorporating a ``teardrop'' effect (see text).}
\end{figure}

\section{Temperature maps of asteroid Steins at the time of the Rosetta flyby}

We now consider the temperature maps of Steins as predicted by our model for the 5 September 2009 Rosetta flyby to discuss the interpretation of the observations performed by the VIRTIS (Coradini et al., 1999; Coradini et al., 2008) and MIRO (Gulkis et al., 2007; Gulkis et al., 2010) instruments. We perform the computation at two different times, 10 minutes before and at closest approach, and for two different combinations of thermal inertia and roughness: (i) $I$=0~JK$^{-1}$m$^{-2}$s$^{-1/2}$ and $\eta$=1.05, and (ii) $I$=100~JK$^{-1}$m$^{-2}$s$^{-1/2}$ and $\eta$=0.79. Figure~8 illustrates the results. These maps are available in digital form on request from the authors. Steins rotates by about 10$^{\circ}$ in 10 minutes inducing only a slight change in the surface temperature distribution. However, the viewing angles are quite different, revealing different regions of the surface. 

Any differences between the two cases (i) and (ii) is not reflected in the maximum temperature, 252~K for 0~JK$^{-1}$m$^{-2}$s$^{-1/2}$ and $\eta$=1.05 and 245~K for 100~JK$^{-1}$m$^{-2}$s$^{-1/2}$ and $\eta$=0.79, but prominently in the temperature distributions. In the case of a null thermal inertia, there is instantaneous adaptation of the surface to the solar insulation and the temperature drops to 0~K in the shadowed regions and night side (Fig.~8, middle panel). This leads to large temperature gradients on the surface close to the terminators, and a maximum day/night amplitude. For a thermal inertia of 100~JK$^{-1}$m$^{-2}$s$^{-1/2}$, the amplitude of the day/night temperature variations is smaller than for 0, and the night side temperature only drops to about 150~K, close to the equator. Moreover, the adapation to solar insulation is not instantaneous as for a null thermal inertia, the highest temperature location being shifted to the afternoon and the temperature rising less rapidly in the morning hemisphere (Fig.~8, bottom panel). These are well known effects that, combined with temperature measurements of the surface obtained by VIRTIS or MIRO, can more tightly constrain the thermal properties of Steins.

The MIRO instrument observed Steins in the millimeter (1.6~mm) and submillimeter (0.53~mm) wavelength ranges (Gulkis et al., 2010). From their measurements, which are difficult to interpret because of pointing uncertainties, Gulkis et al. (2010) derived a constraint on the average emissivity of 0.85-0.9 at 1.6~mm and 0.6-0.7 at 0.53~mm. They also give a range of 450-850~JK$^{-1}$m$^{-2}$s$^{-1/2}$ for the thermal inertia but, as conceded by the authors, MIRO itself provides very little direct constraints on the thermal inertia, and this range is only based on their interpretation of the VIRTIS (unpublished) observations. As confirmed by a discussion with the authors, the MIRO measurements alone cannot rule out a lower thermal inertia of 100-150~JK$^{-1}$m$^{-2}$s$^{-1/2}$ as we found.

The VIRTIS observations give a maximum temperature of about 230~K close to the sub-solar point (Coradini et al., 2008), implying a thermal inertia of 140~JK$^{-1}$m$^{-2}$s$^{-1/2}$ with $\eta$=1.0, consistent with our results. According to Fig.~6 of Gulkis et al. (2010) showing unpublished VIRTIS temperature measurements at different local solar phase angles, the diurnal temperature variation is too small to be compatible with such a thermal inertia, and a larger value is required. In this context, we note that a similar problem was encountered for the nucleus of comet 9P/Tempel~1 (Groussin et al., 2007) and it was concluded that this effect most likely results from a small-scale surface roughness (not taken into account when generating Fig.~8), which tends to increase the flux at larger local solar phase angles compared to a smooth surface (Rozitis and Green, 2010). This interpretation clearly differs from that proposed by Gulkis et al. (2010), but until a full and comprehensive analysis of the VIRTIS observations of Steins is performed and published, we cannot solve this issue.

\begin{figure*} [!t] 
\begin{tabular}{cc}
\hspace{0.5cm} \includegraphics[width=60mm]{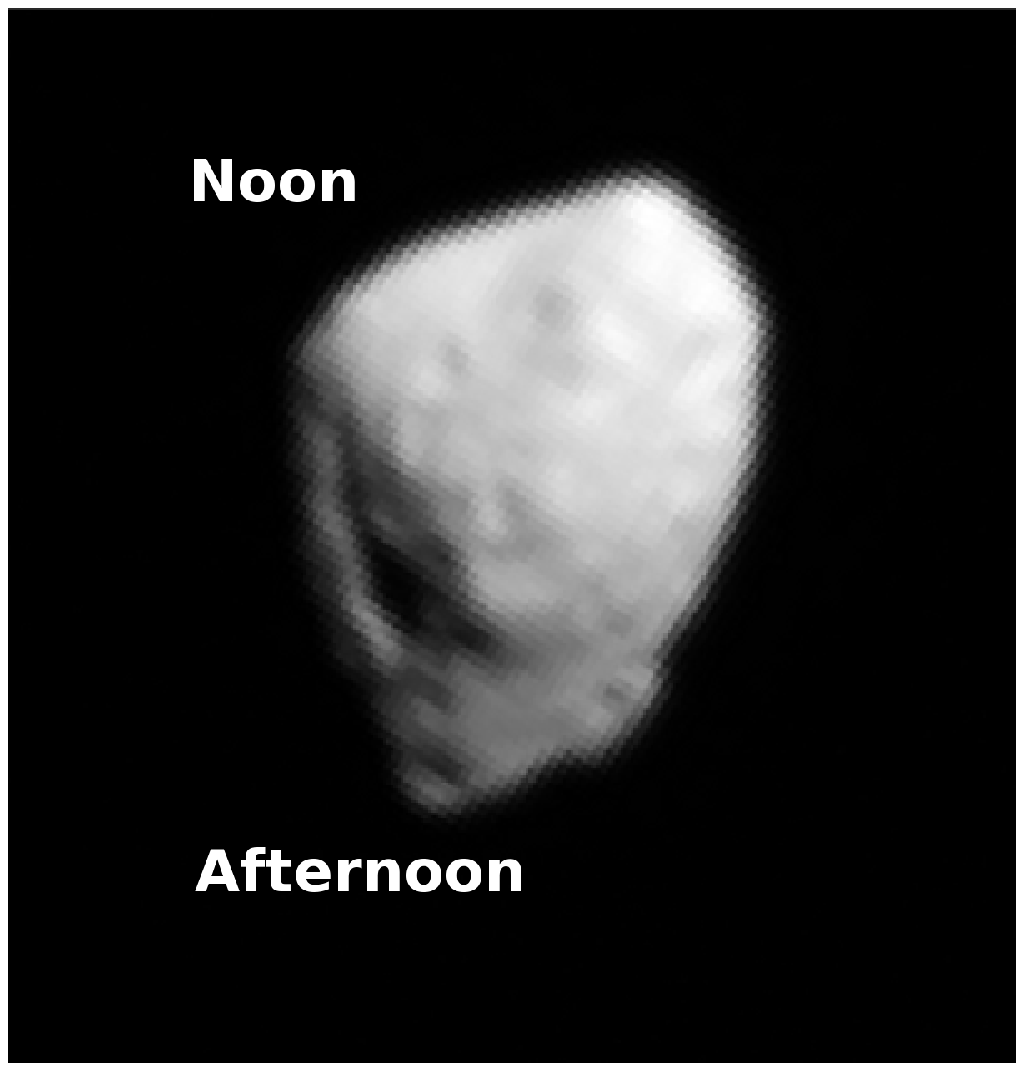} & \hspace{0.2cm} \includegraphics[width=60mm]{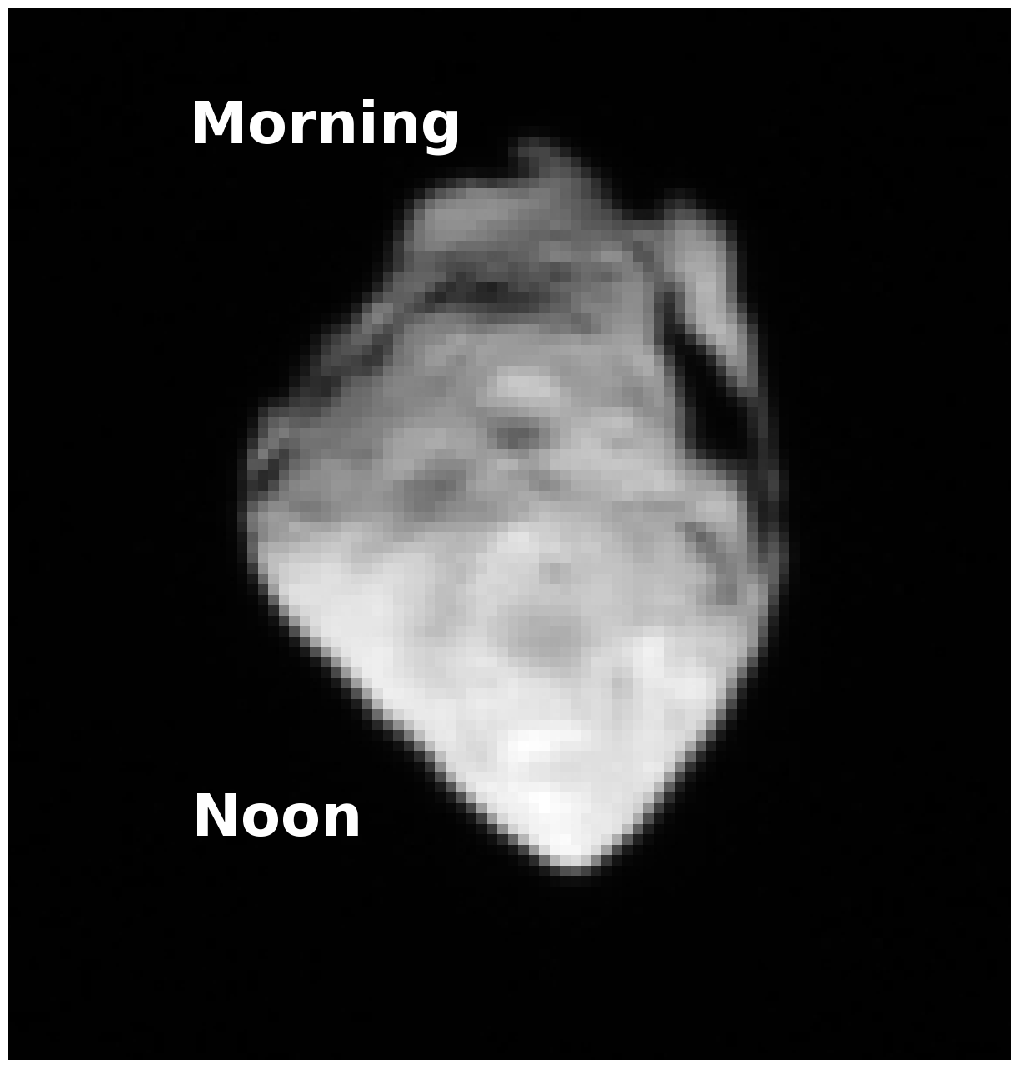} \\
\includegraphics[width=70mm]{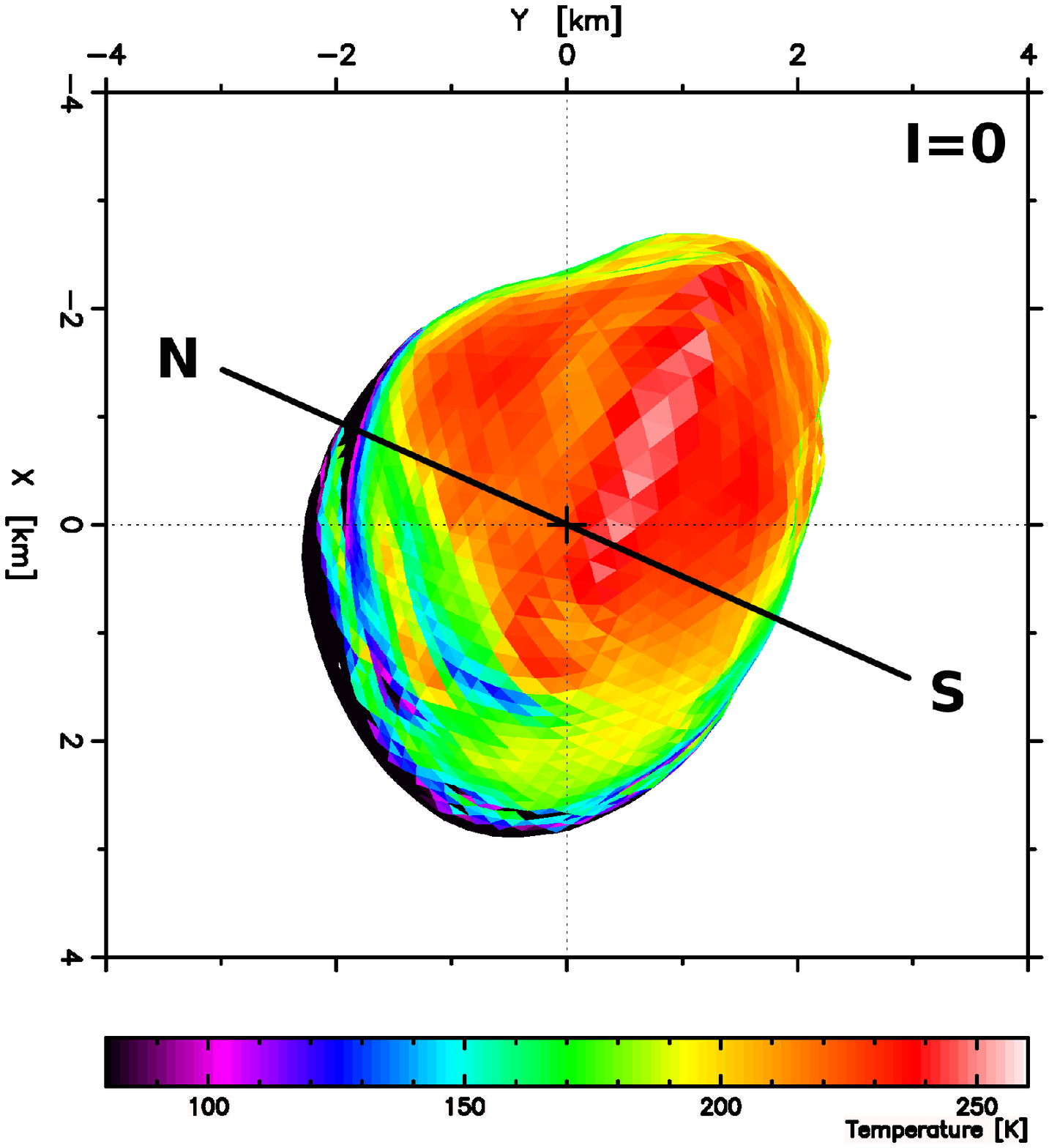} & \includegraphics[width=70mm]{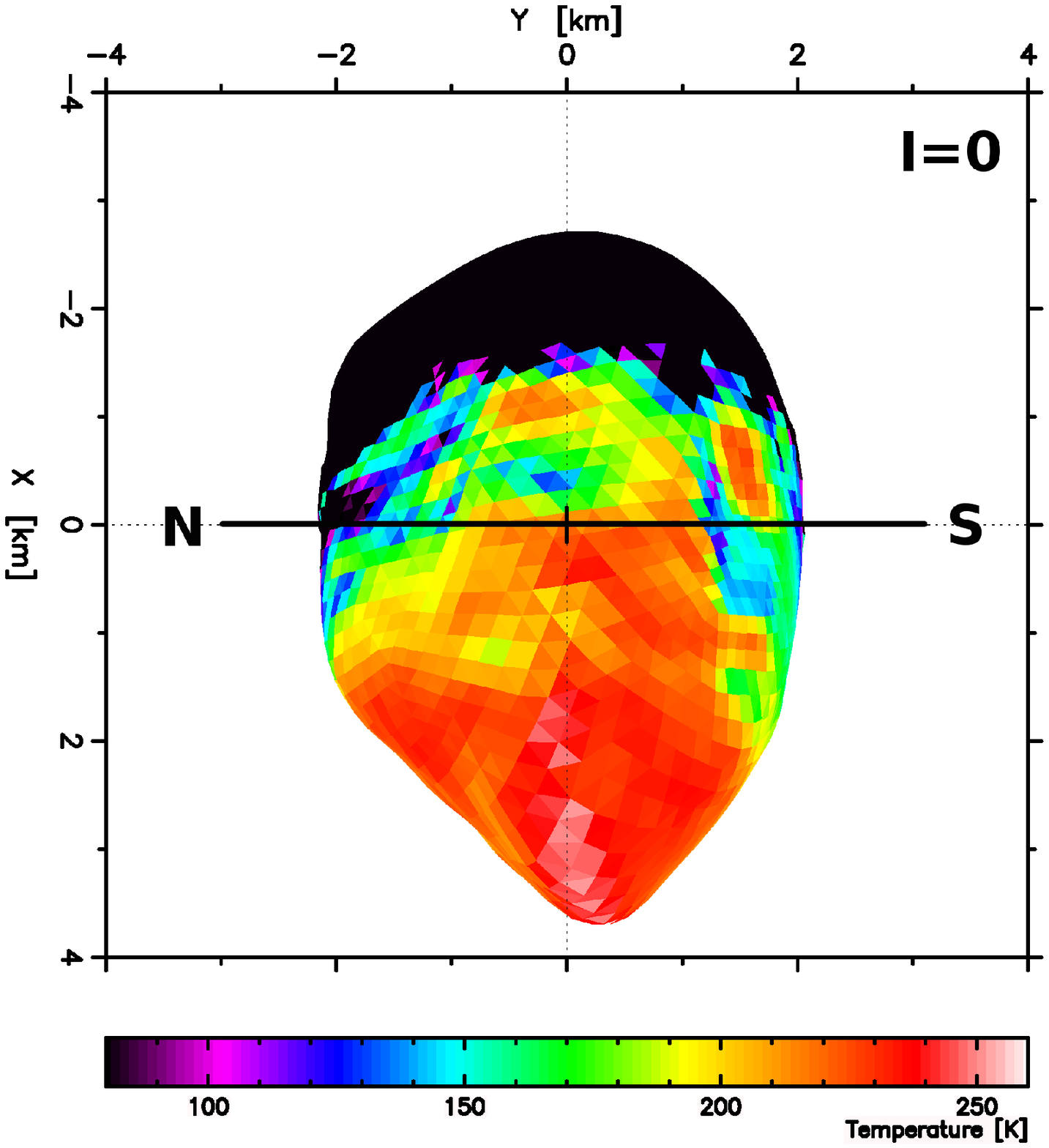} \\
\includegraphics[width=70mm]{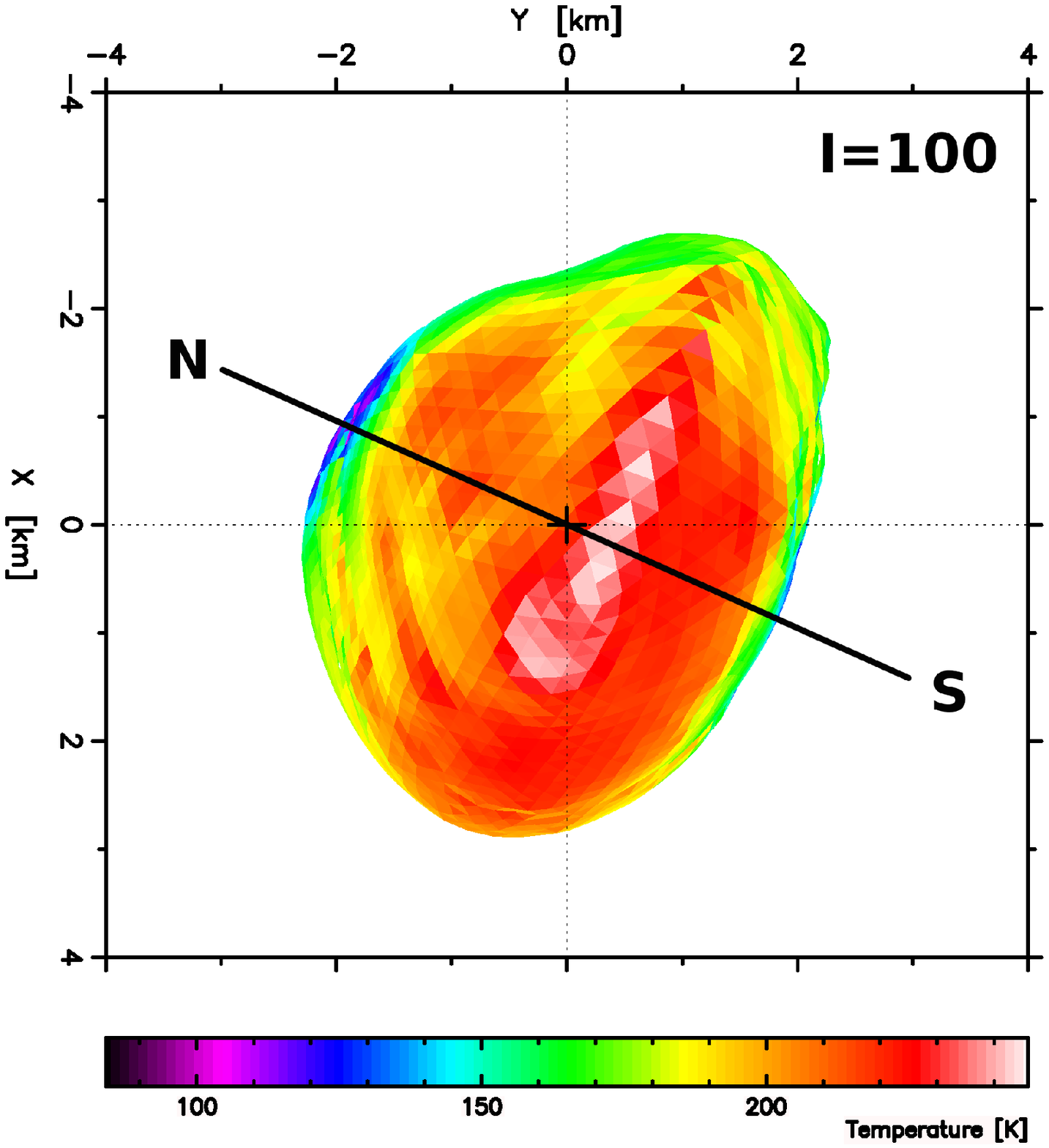} & \includegraphics[width=70mm]{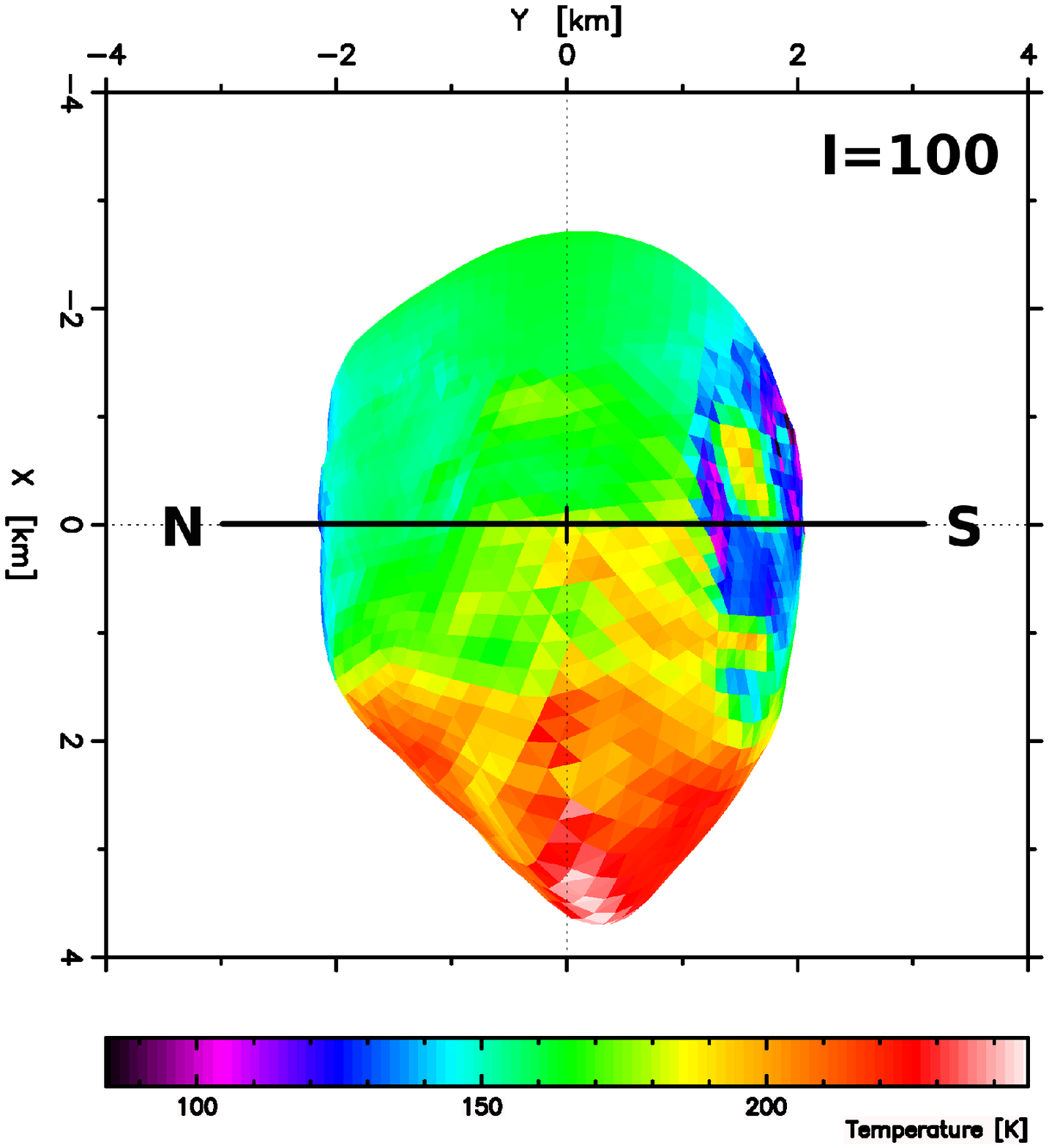} \\
\end{tabular}
\caption{Asteroid Steins as seen from the Rosetta spacecraft during the 5 September 2008 flyby. The left column corresponds to 10 minutes before closest approach, while the right column corresponds to closest approach. The top panel displays images acquired by the OSIRIS camera in the visible. The other two panels display temperature maps: the middle panel corresponds to the case $I$=0~JK$^{-1}$m$^{-2}$s$^{-1/2}$ and $\eta$=1.05, and the bottom panel to $I$=100~JK$^{-1}$m$^{-2}$s$^{-1/2}$ and $\eta$=0.79.}
\end{figure*}

\section{Conclusions}

We acquired new thermal infrared spectra of asteroid Steins with the SST in March 2008 and have interpreted these data, in addition to data obtained in November 2005, using the shape model and photometric properties inferred from the data of the Rosetta flyby of 5 September 2008.
The two datasets are consistent with a thermal inertia of 100$\pm$50~JK$^{-1}$m$^{-2}$s$^{-1/2}$ and a beaming factor (roughness) of 0.7-1.0 (Table~2).
This range of thermal inertia is barely compatible with the constraint $I<$100~JK$^{-1}$m$^{-2}$s$^{-1/2}$ obtained by Delb\'o et al. (2007) from their compilation of main-belt asteroids (MBAs) larger than 100\ts km, but consistent with the results $I<$400~JK$^{-1}$m$^{-2}$s$^{-1/2}$ (Nesvorn\'y \& Bottke, 2004) and $I<$200~JK$^{-1}$m$^{-2}$s$^{-1/2}$ (Harris et al., 2009) for the Karin family, whose mean size of about 4\ts km is comparable to that of Steins.
On the basis of the thermal properties of 17 MBAs in the size range 25-1000\ts km, Delb\'o et al. (2007) and Delb\'o \& Tanga (2009) suggested that the thermal inertia increases with decreasing size. 
Our results for Steins do not confirm this putative correlation for kilometer-size MBAs as it would have implied a thermal inertia of several thousands JK$^{-1}$m$^{-2}$s$^{-1/2}$ for Steins based on the correlation law of Delb\'o \& Tanga (2009).

We confirm that the infrared emissivity of Steins is consistent with an enstatite composition, in agreement with the results of Barucci et al. (2008).

The November 2005 SST thermal light curve is more accurately interpreted by assuming inhomogeneities in the thermal properties of the surface, or more precisely two different regions with different roughness. 
A small change in the beaming factor of about 0.1 is required between the two different regions, which is compatible with observations of other small bodies, for example the nucleus of comet 9P/Tempel~1 (Li et al., 2007; Spjuth et al., 2011).

Overall, our results emphasize that the shape model is important to an accurate determination of the thermal inertia and roughness.

However, uncertainties remain, in particular for the spatial variations in the thermal properties, which requires in-situ measurements of the surface temperature distribution, as illustrated for the nucleus of comet 9P/Tempel~1 (Groussin et al. 2007). 
We hope to make progress in this direction with the flybys of asteroid 21~Lutetia in July 2010 by Rosetta, of comet 103P/Hartley~2 in November 2010 by EPOXI, and even more promising with the Rosetta rendez-vous with comet 67P/Churyumov-Gerasimenko in 2014.

\begin{acknowledgements}
This work is based on observations made with the Spitzer Space Telescope, which is operated by the Jet Propulsion Laboratory, California Institute of Technology under a contract with NASA.
We  thank the SST ground system personnel for their prompt and efficient scheduling of the observations.  
\end{acknowledgements}

\end{document}